%% file: beads.tex
\documentclass[conference]{IEEEtran}

\usepackage{blindtext, graphicx, todonotes, verbatim, url, amsmath, amssymb,
subfigure, balance}
\usepackage{algorithm}
\usepackage{algpseudocode}
\usepackage{paralist}
\usepackage{amsmath, amssymb, amsthm, framed, graphics}
\usepackage{fancybox,fancyvrb}
\usepackage{cite}

\newcommand{\pint}{{\em pInt}}
\newcommand{\pints}{{\em pInt} messages}
\newcommand{\erase}{{\sf erase}}

\begin{document}

\newtheorem{definition}{Definition}

\title{BEAD: Best Effort Autonomous Deletion in Content-Centric Networking}

\author{\IEEEauthorblockN{Cesar Ghali \quad\quad Gene Tsudik \quad\quad
Christopher A. Wood}
\IEEEauthorblockA{University of California, Irvine\\
Email: \{cghali, gene.tsudik, woodc1\}@uci.edu}}

% make the title area
\maketitle
\pagestyle{plain}

\begin{abstract}
A core feature of Content-Centric Networking (CCN) is opportunistic
content caching in routers. It enables routers to satisfy content requests
with in-network cached copies, thereby reducing
bandwidth utilization, decreasing congestion, and improving
overall content retrieval latency.

One major drawback of in-network caching is that content producers have no
knowledge about where their content is stored. This is problematic if a
producer wishes to delete its content. In this paper, we show how to address
this problem with a protocol called BEAD (Best-Effort Autonomous Deletion).
BEAD achieves content deletion via small and secure packets that resemble
current CCN messages.
We discuss several methods of routing BEAD messages from producers to caching
routers with varying levels of network overhead and efficacy. We assess
BEAD performance via simulations and provide a detailed analysis of its properties.
\end{abstract}

\IEEEpeerreviewmaketitle

% meat and potatoes
\input{01-intro}
\input{02-overview}
\input{07-related}
\input{03-initial}
\input{04-analysis}
\input{05-experiments}
\input{06-discussion}
\input{08-conclusion}

\balance

\bibliographystyle{IEEEtran}
\bibliography{IEEEabrv,references}

% that's all folks
\end{document}

%% file: 01-intro.tex
\section{Introduction}
Content-Centric Networking (CCN) is a relatively recent internetworking paradigm
touted as an alternative current IP-based Internet architecture.
While IP traffic consists of packets between communicating end-points, CCN traffic is
comprised of explicit requests for, and responses to, named content objects.

An important features of name-based content retrieval is decoupling of content from
its producer. This enables more natural content distribution by allowing routers to
opportunistically cache content \emph{within the network}. Cached content can be
returned in response to future requests, which are called {\em interests}. This reduces
the need to forward interests to content producers, thus lowering network congestion and
content retrieval latency.

However, router caches are not mandatory in CCN. In some cases, caching
content might not be beneficial, e.g., for routers with high content
processing speeds, since high arrival rates translate to less time spent in cache.
If this cache lifetime is very short, the probability of cache misses
increases and cache's utility decreases proportionally. Indeed, some prior literature shows
(via simulations and experiments) that caching at the edges of the internetwork, i.e.,
at consumer-facing routers, is most beneficial and more cost-effective than doing so in
the core, i.e., in transit routers \cite{dabirmoghaddam2014understanding}.

To help caching routers determine the lifetime of cached content, the latter includes
an optional {\tt ExpiryTime} field. Routers are expected to flush content once this
time elapses. However, a router can choose to keep content cached beyond its lifetime.
Lifetime of content in a particular router's cache depends entirely upon
that router's implementation and policy. This uncertainty (or freedom) means that content
may linger in the network for a very long time.

%These caches are managed with common algorithms such
%as Least Recently Used (LRU) and Least Frequently Used (LFU) \cite{abdullahi2015survey}.

One notable drawback of this libertarian approach to caching is that some
content may need to be deleted \emph{before} {\tt ExpiryTime} elapses.
Consider content that frequently (yet sporadically) evolves over time,
e.g., news articles. The appearance of breaking-news articles is
unscheduled. As situations develop, updates and corrections to
the content occur at unpredictable times. Such updates supersede
previously distributed content by rendering it stale. Thus, in this
case, producers need a way to remove old content. Another example is content
(that has released and subsequently cached) which contains erroneous information.
As errors are detected and corrected, a producer needs to flush the
incorrect older version.

The deletion problem occurs because {\tt ExpiryTime} is the only way
for a producer to communicate {\em anticipated} content lifetime to
the network. However, a producer can not change its mind after content has
been published and distributed. Thus, there is a need for a safety
mechanism for in-network content deletion.  In this paper, we design a
protocol -- Best-Effort and Autonomous Deletion (BEAD) -- that mitigates
this problem. In the process, we encounter and address several challenges,
including efficacy, efficiency and security. We also experimentally
assess the proposed BEAD protocol.

%%%CESAR: this part is commented out by Gene.
%
%Our contributions are three-fold:
%
%Specifically, producers should
%be able to send deletion requests to routers that might have cached offending
%content. Routers would verify such requests and delete corresponding content,
%if cached. Clearly, deletion requests should be secure in order to prevent malicious content
%flushing.%
%\begin{compactitem}
%\item A lightweight protocol (BEAD)  for authenticated content deletion.
%\item  route content deletion requests from
%producers to consumers.
%\item Specification of the complete BEAD protocol with corresponding analysis and
%experimental assessment.
%\end{itemize}
%
The rest of this paper is organized as follows. Section \ref{sec:overview} overviews CCN. Related
work is summarized in Section \ref{sec:related}. Section \ref{sec:reqs} presents minimal
requirements for content deletion. Sections \ref{sec:auth} and \ref{sec:routing} describe
authentication and routing of deletion requests in BEAD, respectively. The BEAD protocol
is analyzed in Section \ref{sec:analysis} and its performance is assessed in Section \ref{sec:simulations}.
The paper ends with a discussion of BEAD optimizations and practical
factors in Section \ref{sec:monotizing}. Future work is summarized in Section \ref{sec:conclusion}.

%%%CHRIS: I just removed this outright. It is a type of stale content and not really
% motivating or useful since one wouldn't cache content that needs to be recycled
% this long.
%It might be trivially imagined that BEAD is suitable for deleting content
%under access control. Consider the scenario where a
%consumer $Cr$ has access to a sensitive content object $C$ that is cached
%somewhere in the network. If $Cr$'s access is revoked \emph{before} $C$ is
%evicted from the cache, $Cr$ could possibly still retrieve said content by
%issuing a request for $C$ as normal. This is because routers \emph{do not} perform any
%form of authorization checks for content before responding from the cache; they
%are completely oblivious to the nature (e.g., sensitivity) of content they
%process. However, the fact that BEAD is a best-effort deletion mechanism, it
%might not be suitable for deleting content under access control, especially
%if the content contains sensitive and confidential information. In this case,
%it is better not to cache such content in routers.

%% file: 02-overview.tex
\section{CCN Overview} \label{sec:overview}
This section provides an overview of the CCN architecture and protocol,
according to the most recent specification \cite{mosko2015ccnx}. Given
familiarity with CCN, it can be skipped without loss of continuity.

Unlike IP, which focuses on addressable end-hosts, CCN emphasizes named and
addressable content. A consumer issues
a request, called an {\em interest}, specifying the name of desired
content. CCN names are structured simular to URIs. For example, a particular
content produced by the NSA might be named: \path{lci:/us/gov/DoD/NSA/Snowden-Diary}. 
% Depending on the context, we use the notation $N$ to refer to the name of an
%interest or content object. 
An interest for a particular content $N$ is routed towards an authoritative producer
for the content based on $N$ itself. In CCN, both interest and
content messages have general-purpose {\tt Payload} fields. Consumers can
use an interest's {\tt Payload} field to \emph{push} information to
producers, while producers use a content's {\tt Payload} field to
carry actual application data.

As an interest traverses the network, each router determines
if a copy of requested content is cached in its Content Store (CS).
%\footnote{A cache 
%entry contains the name, cryptographic hash digest, and the actual payload
%of the content.}
If a cache hit occurs, the router satisfies the interest by sending the matching
content on the interface on which the interest arrived. Otherwise, the
router (1) records some state derived from the interest in its Pending Interest
Table (PIT) in order to provide a backwards path for the future content, and (2)
forwards the interest to the next hop(s) specified in its Forwarding Information
Base (FIB). State retained in the PIT contains the content name and the
interface(s) on which interests for that name have been received. A FIB is
a routing table that maps hierarchical name prefixes to outbound interfaces.
Longest-Prefix Matching (LPM) is used to determine the matching FIB entry.

A router $R$ can collapse multiple interests into the same PIT entry
whenever:
\begin{compactenum}
\item $R$ receives an interest for name $N$
\item $R$ does not have content $N$ in its cache
\item $R$'s PIT already contains an entry for $N$
\end{compactenum}
When interest collapsing occurs, $R$ only records the interface on which the new
interest arrived and drops that interest. Whenever requested content arrives, $R$
forwards it on all interfaces listed in the corresponding
PIT entry. Afterwards, the PIT entry is flushed.

If no router can finds a cached copy of requested content in its cache, the interest
eventually reaches the producer that responds with the matching content, if possible. 
If the producer can not provide it (e.g., content does not exist) a NACK is generated 
\cite{mosko2015ccnx,compagnonack}. As content  traverses the
reverse path to the consumer, routers may choose to cache it in anticipation of
future requests. As mentioned earlier, each content includes a producer-set {\tt ExpiryTime} field. 
This value is content- and application-specific. However, each router
can use any cache management algorithm, e.g., LRU or LFU. 
%If content is not evicted from the cache by the management algorithm,
%then it \emph{should} be evicted when its {\tt ExpiryTime} expires.

%%%CHRIS: I removed all of this security stuff. It's not relevant to the paper.
%Beyond the pull-based communication model that guarantees symmetric interest and content
%object flow, content-centric traffic in CCN has strong security properties. Notably,
%security is coupled to content rather than its distribution channel. All
%sensitive content must therefore be encrypted in a meaningful way so as to ensure
%confidentiality.\footnote{This is not the only way to ensure confidentiality, see
%\cite{ghali2015interest} for an alternative approach based on encrypting and authenticating interests.}
%Content integrity and origin authenticity are ensured via digital signatures
%generated by content producers. This is not a requirement, though, as the
%authenticity of content may be ascertained by checking the cryptographic hash
%digest of the content that is returned. CCN leverages Manifests
%\cite{kurihara2015encryption} to publish hash digests so that self-certifying
%interest messages -- interests that contain the hash digest of the desired
%content object -- can be issued and extensive signature verification can be
%avoided at routers. See \cite{kurihara2015encryption,ghali2014network} for more
%details about this content retrieval strategy.

%% file: 07-related.tex
\section{Related Work} \label{sec:related}
Lack of on-demand content deletion is a well-known problem in CCN\cite{mauri2013distributing,yuendorsement,ghali2015interest,ghali2014network,yu2015schematizing,wood2014flexible}.
The problem of {\em unsafe replicas} or stale content in CCN was first considered in \cite{angiusdrop}.
Analytical and experimental assessment showed that: ``...the more frequently
content is requested the higher is the chance of one request ending up in
between a revocation and the eviction [of the stale key].'' The proposed solution relies on
a monotonically decreasing cache lifetime enforced by cooperating routers.
This does not allow a producer to change the lifetime
after content is published; it only seeks to minimize the time window when
stale or unsafe replicas can be accessed.

\cite{mauri2013distributing} proposed a mechanism to implement
revocation of content without input from the consumer. The proposed approach
uses the ccnx-sync protocol to perform OCSP-like \cite{ocsp} synchronization of key data, i.e.,
determine content that has been revoked. This requires proactive behavior
by each participating repository. % and is not distributed like BEAD.
\cite{yuendorsement} suggests using ChronoSync \cite{zhu2013let} to
synchronize revoked key endorsements among group members. Revocation,
however, is not the same as cache deletion. Revoked content, if still cached, can be
inadvertently accessed by malicious or benign consumers.

\cite{ishiyama2012effectiveness} discussed a new caching technique
allowing routers to proactively share content with downstream peers which
did explicitly request  that content. The suggested multicast forwarding strategy
serves to increase the number of replicas in the network. However, unsolicited
content objects can be seen as a form of attack similar to
cache poisoning \cite{ghali2014network}.

The concept of cost-aware caching in CCN was introduced in
\cite{agyapong2012economic,araldo2014cost,wang2012incentivizing,araldocost,Suksomboon2012}.
Various economic incentives for ISPs and ASs to cache content on behalf of
producers have been explored. Cost-aware routers that
cache based on popularity and economic incentives are studied in \cite{araldo2014design}.
In general, the economic problem of supporting prioritized caching in the network
is addressed without any attention to the inverse problem: how is content removed
from caches? 

%%%CHRIS: if space permitted, add an example about the difference above^^

%% file: 03-initial.tex
\section{Protocol Requirements} \label{sec:reqs}
Our motivation stems from the need to remove stale or erroneous
content from the network, i.e., from routers' caches. One intuitive way of doing this
is through the use of versioning, whereby the content naming format
includes a component that explicitly reflects the current version. For example,
the content of BBC's World News web-page could be named:
\path{lci:/bbc/news/world/v2.4}. One immediate drawback to this approach is
that consumers cannot be expected to know the latest version in order
to form such a name. Alternatively, timestamps could be used.
In that case, the same BBC page could be named
\path{lci:/bbc/news/world/1449187200}.\footnote{1449187200 is 12/04/2015 at 12:00am UTC.}
%%%CHRIS: this is not true. how is it unclear? just put the current time in the interest.
%%%       I think the problem lies in caching -- two consumers will not request at the same
%%%       time to benefit caching, so why bother caching at all?
However, it is unclear how consumers would determine such
timestamps, without which they can not request content.

The main problem with versioning and timestamps is that they can not handle
unpredictable content updates. In current CCN design, producers are
oblivious to where and for how long their content is
stored in the network. Although this opportunistic
caching is one of the biggest CCN advantages, it
greatly complicates deletion of stale content.
We believe that, in order to address the problem, producers need:
\begin{compactenum}
    \item A way to communicate a single deletion request to all routers that  might have
    cached offending content.
    \item A way to efficiently secure deletion requests (allowing routers to quickly
    authenticate them) while avoiding trivial Denial of Service (DoS) attacks.
\end{compactenum}
The first requirement is reminiscent of IP traceback --  a class of techniques
for identifying the original source of a (usually malicious) packets. In the
context of IP, this is often framed as a mechanism to help stop Denial of Service (DoS)
attacks. In the context of this paper, the goal is to learn {\em where} content was
previously forwarded so that deletion requests can be routed along the same paths. These
paths correspond to the original sources of interests for that content. Thus, ideas from IP
traceback based on packet logging (e.g., \cite{snoeren2001hash}) and (deterministic or
probabilistic) packet marking (e.g., \cite{goodrich2002efficient,belenky2003ip})
influence the design and forwarding strategies of BEAD messages.

We now present solutions to these requirements that are incorporated into the
BEAD protocol.

\section{Authenticating Deletion Requests} \label{sec:auth}
Producers must prove content ownership to routers that receive deletion requests.
Otherwise, an adversary can impersonate a producer and induce content deletion,
resulting in a form of DoS. One way to attain authentication is is by a producer-generated
digital signature on each deletion request. However, besides being inefficient,
forcing routers to verify signatures on deletion requests can be parlayed into denial-of-service
(DoS) attacks \cite{gasti2013and,ghali2014network}. % based on verifying said signatures.
Moreover, it involves public key retrieval, certificate handling and other messy
(for routers) issues.

Our approach uses a light-weight token that proves content ownership.
When a producer $P$ creates a content object $C$, it generates
a random $\lambda$-bit string $x_C$, called the {\em deletion token}. $P$ then computes
the digest of this token using a suitable cryptographic hash
function\footnote{Suitable hash functions include those with
pre-image resistance, which means that, given $y$, it is difficult
to find an $x$ such that $y = H(x)$.} -- $y_C = H(x_C)$ -- and
includes $y_C$ in $C$. Later, if and when $P$ wishes to
delete $C$ from the network, it includes $x_C$ in the deletion
request. (We assume that $P$ can route these requests to any
router caching $C$.) Upon receipt, each $R$ verifies that $y_C$ (cached
alongside the content) matches $H(x_C)$. If so, $R$ knows
that $P$ must have issued the request and deletes $C$ from
the cache.\footnote{This is due to the randomness of $x_C$ and the
collision-resistance of $H(\cdot)$.}

\section{Routing Deletion Requests} \label{sec:routing}
The remaining (though major) issue is how to route deletion requests from the producer
to each caching router. This can be viewed as a multicast problem where
producers must distribute a message (deletion request) to only a subset of
nodes which could have cached the content.

%%GTS: this is not good... Needs to be in a better form
%%CHRIS: how is it not good? I don't see the problem.
Let $I\/nt[N]$ and $C[N]$ be the interest and content messages with the name $N$.
The hash digest of $C[N]$ is a $\lambda$-bit string $d$, i.e., $d = H(C[N])$.
Let $E[N, d]$ be a deletion request for content with the name $N$ and hash digest $d$.
Let $\mathbb{R}_N$ be the set of routers which cached $C[N]$. Finally, let the FIB
of router $R \in \mathbb{R}_N$ be FIB$^{R}$.

For the rest of the paper, we use the terms
\erase, \erase\ message and \erase\ messages to refer to deletion requests.
Also, we assume that \erase\ messages are authenticated using the method
described in Section \ref{sec:auth}.

\subsection{Flooding}
We begin by considering the simplest approach: reverse-path controlled flooding
\cite{baker2004ingress} of deletion requests. When $R \in \mathbb{R}_N$ receives
$E[N,d]$, it forwards it on all interfaces except those which have a matching FIB
entry, as shown in Algorithm \ref{alg:flood}.

\begin{algorithm}[t]
\caption{\erase-Flood} \label{alg:flood}
\begin{algorithmic}[1]
\State {\bf Input:} $R \in \mathbb{R}_N$ and $E[N,d]$
\State $\mathsf{faceset} :=$ FIB$^{R}.\mathsf{Lookup}(N)$
\For {$\mathsf{face} \notin \mathsf{faceset}$}
    \State {\sf Forward} $E[N,d]$ to $\mathsf{face}$
\EndFor
\end{algorithmic}
\end{algorithm}

Flooding offers some advantages, the most important of which
is the ability to reach edges of the network even if routers on the
producer-to-consumers paths no longer cache the content to be deleted.
This is important since routers do not cache content uniformly and
some may not even have caches.
On the negative side, the amount of traffic generated from a single deletion
request is very high and most deletion requests would be
forwarded to routers that never even had the target content.

\subsection{Forwarder Histories for Content Traceback} \label{sec:fw_histories}
In the optimal case, routers would only forward \erase\ messages on interfaces to which
the referenced content had been previously forwarded. In other words, \erase\ messages
should only be forwarded along the content distribution spanning tree where the producer
is the root and leaves are the consumers who requested the content.
One way to forward \erase\ messages along the edges of this tree
is for each router $R \in \mathbb{R}_N$ to maintain some forwarding history of $C[N]$.
There are several places where this history can be stored, including: (1) in
%%%CESAR: comment the following two lines and uncomment the ones after if we delete
%%%       interest marking.
the cache where $C[N]$ is stored, (2) in a forwarding log (similar to \cite{snoeren2001hash},
as a form of IP traceback) at each routers, and (3) in the packets themselves.
%the cache as $C[N]$ is stored, and (2) in a forwarding log (as suggested in \cite{snoeren2001hash}
%as a form of IP traceback) at each routers.
%%%%%%%%%%%%%%%%%%%%%%%%%%%%%%%%%%%%%%%%%%%%%%
In each case, historical information constitutes a form of traceback
that allows routers to identify where content was previously forwarded.
We now describe each approach in more detail.

\subsubsection{In-Cache Forwarding Histories}
When a router caches $C[N]$ it can also remember the downstream interfaces
where the cached copy was forwarded. We denote the set of these
interfaces as $\mathbb{F}_N$. When a router
receives an interest $I\/nt[N]$ on interface $F_i$, it responds with $C[N]$
and adds $F_i$ to $\mathbb{F}_N$. For a router
with $K$ interfaces, this additional state costs $\mathcal{O}(K)$
bits per cache entry. When a router caching $C[N]$ receives
$E[N,d]$, it forwards it on all interfaces in $\mathbb{F}_N$.

In-cache forwarding histories are only effective for routers that have large caches,
since the lifetime of forwarding information is bound to the
lifetime of cache entries, which can be small or even zero (if a
router has no cache at all). Since a forwarding history $\mathbb{F}_N$ is deleted
whenever $C[N]$ is flushed from the cache, this can lead to a future
$E[N,d]$ not being forwarded to downstream routers which might still cache $C[N]$.

\subsubsection{Local Forwarding Logs}
\begin{figure}
\centering
\includegraphics[scale=0.5]{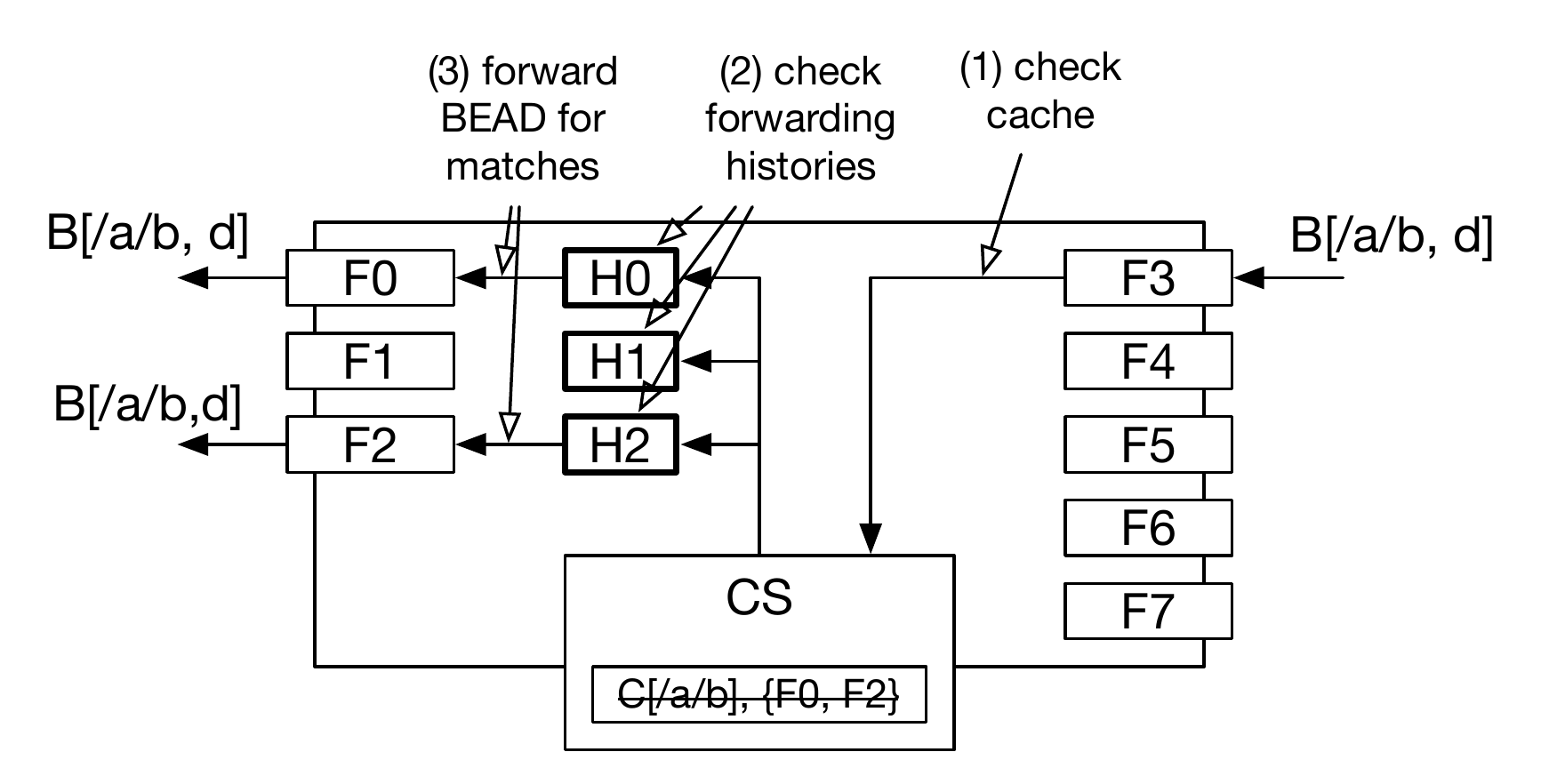}
\caption{$E[N,d]$ forwarding strategy based on per-interface forwarding
histories.}
\label{fig:lookup2}
\end{figure}

Long-term packet logs have their roots in IP traceback techniques from the early
2000-s, e.g., \cite{stone2000centertrack,snoeren2001hash}. The problem here is similar: routers need
long-term histories of packets (content) that were previously processed and
forwarded. In this context, a history is a set-like data structure that allows
content objects to be inserted and then later queried for membership. There are
two types of histories: {\em lossless} and {\em lossy}. The former
always return ``yes'' for content objects that have previously
been inserted. In contrast, a lossy history might return false positives
or negatives. Routers use these structures by associating one
history to each interface. When a router receives $E[N, d]$ and $C[N]$ is not cached,
it forwards $E[N,d]$ on each interface for which the corresponding forwarding
interface history has a record of $C[N]$, i.e., all histories for which
membership query returns ``yes''. This procedure is outlined in
Figure \ref{fig:lookup2}.\footnote{Similar to the flooding algorithm,
this check is not performed for interfaces via which the content producer
can be reached.}

We now describe how to implement lossless and lossy histories that vary in
their computation and memory requirements.

\noindent
{\bf Lossless Forwarder Histories}
require  a unique identifier to be kept after a content
object has been forwarded. We assume that content hash digest $d$ serves as such an identifier
(with collision probability negligible in $\lambda$). Implementing this type of
forwarder history can be done trivially with a hash set $HS_R$ as follows: to insert a
content object into the history, compute and store $d$ in $HS_R$. To query the
history, return ``yes'' if $d \in HS_R$ and ``no'' otherwise.
Insertion and lookup each require constant time.

\noindent
{\bf Lossy Forwarder Histories}
are intended to store historical information
in memory-constrained systems at the cost of false positives
and false negatives. Similar to the SPIE traceback mechanism \cite{snoeren2001hash}, we use
Bloom Filters (BFs) \cite{bloom1970space} to implement lossy forwarder
histories. BFs enable the required probabilistic set membership queries.

The choice of BF properties, e.g., size and hash functions, impacts
efficacy of this technique. Filters that saturate too quickly result
in high false positive rates. If all interface filters become saturated
then \erase\ is effectively broadcasted. Therefore, it is important to
eventually remove stale elements from filters. Unfortunately, a regular
BF does not provide element removal. However, so-called Counting Bloom Filters (CBFs)
\cite{fan2000summary} support set membership queries with removal. Instead of using bits
to indicate set membership, CBFs use counters. When loading an element into CBF, the
counters corresponding to the output of the hash functions are increased by one.
Consequently, removing an elements is done by decrementing the same
counters. The problem with CBFs is that one must know the element to delete.
Since routers would discard content after inserting them into these filters\footnote{
This is because content is only added to histories upon its removal
from the cache.}, they have no way of knowing what content is in the filter, and thus what
elements to eventually delete. Their only recourse is to remove
elements by decrementing counters at random. Intuitively, a router would
delete random elements from the filter (the history) at a frequency which reflects the average
{\tt ExpiryTime} of received content. This can increase the false negative probability
and reduce the possibility of delivering \erase\ messages to their corresponding destination.

Variants of the CBF, such as Time-Decaying (TDBFs)
\cite{zhang2008detecting,koloniari2011one} and Stable (SBFs)
\cite{deng2006approximately} BFs can also be used. TDBFs have the property that
elements are slowly removed from the filter over time, thereby keeping the rate
of false positives minimized. However, the natural decay property may lead to
false negatives. SBFs on the other hand are dynamically self-resized to keep the false
probabilities minimized. Similar to CBFs and TDBFs, SBFs also suffer from
false negatives.

\subsubsection{Interest Marking for Content Traceback} \label{sec:interest_marking}
Packet marking is a standard technique in IP traceback techiques \cite{goodrich2002efficient}.
In the context of this work, marking is performed on interests to indicate
sources of content requests. This information can be later used
to learn the interface to which an \erase\ needs to be sent.
Specifically, \erase\ messages can carry this marking information in order for routers
to identify the appropriate downstream interfaces without storing any local state.

One trivial marking method is to append the arrival interface to
each interest. Specifically, when $R$ receives $I\/nt[N]$ on face $F_i$,
$R$ prepends $(R, F_i)$ to a list contained in
the header of the interest. Producers record these traces upon receipt.
In the event that an \erase\ needs to be generated, $P$ includes the
trace in the \erase\ and forwards it on the appropriate downstream interface.
When $R$ receives an \erase\ with a trace it pops the last element
$(R,F_i)$ off the trace list and forwards it on the specified interface $F_i$.

This technique distributes the forwarding history among messages
in the network. As a consequence, this information must be secure. To illustrate
this requirement, assume router $R_i$ receives $E[N,d]$ with the sequence of hops
\begin{align*}
[(R_i, F_i),(R_{i-1},F_{i-1}),\dots,(R_2, F_2),(R_1, F_1)]
\end{align*}
from interface $F_{i+1}$. $R_i$ needs a way to securely guarantee that the
tuple $(R_i, F_i)$ was previously prepended, by itself, to the subsequence:
\begin{align*}
[(R_{i-1},F_{i-1}),\dots,(R_2, F_2),(R_1, F_1)].
\end{align*}
Otherwise, malicious entities can forge unsolicited \erase\ messages with
apparently correct routing sequences. Alternatively, one can modify existing sequences
in \erase\ messages to prevent them from being routed towards their destination.

One way of authenticating hop-sequence traces is for
$R_i$ can compute a Message Authentication Code (MAC)
\cite{krawczyk1997hmac,gutmann2012using} tag $t_i$ over the (relevant) interest details, e.g., the
name and previously present traces in the hop-sequence. $R_i$ then adds the
tuple $(R_i, F_i, t_i)$ to the interest before forwarding it. Since \erase\ messages
carry the name of the content to be deleted, each router will be able to verify
its pre-computed tag before forwarding \erase\ messages downstream.
Since routers compute and verify tags locally,
a key management and distribution protocol is not required. We do, however,
assume that routers are able to generate and maintain cryptographic keys
of sufficient length necessary for MAC computation. As an added feature,
hop-sequence information can also be used for detecting both interest and
\erase\ loops \cite{garcia2015enabling}.

Although this technique of marking interest is effective to deliver \erase\ messages to all
routers on the path between consumers and producers, it has several drawbacks.
One of this is that interest traces received by producers need to be stored so that
they can be included in \erase\ messages. This due to the fact that (1) each trace corresponds
to only one path in the network, and (2) interests issued by multiple consumers
are most likely to traverse different paths to the producer.
Producers can attempt to compile all collected traces in a data structure forming a spanning
tree. This structure would be included in \erase\ message headers,
allowing routers to forwarder \erase\ messages correctly. The main disadvantage of this approach
is that the size of the data structure grows linearly with the number of consumers
and is most likely to be greater than average allowed MTU. This means that \erase\
messages will be fragmented (and possibly re-fragmented), and hop-by-hop reassembly is not
avoidable. Another alternative is for producers to send multiple \erase\ messages one for each
set of traces correlated to a hop-sequence. In Section \ref{sec:analysis}, we compare and
evaluate the performance and resource consumption of these two techniques.

%% file: 04-analysis.tex
\section{Analysis} \label{sec:analysis}
In this section we assess some routing strategies
for \erase\ messages. % from producers to caching routers.
Let $n_t^R$ be the total number of
content objects forwarded by $R$ at time $t$ and let $\mu_F^R$ be $R$'s
content forwarding rate. Note that $n_t^R$ grows
monotonically as a function of $\mu_F^R$.

\subsection{Flooding Analysis}
Recall that the reverse path flooding algorithm works by only sending broadcast
messages to interfaces through which the \emph{producer} is not reachable.
Though very effective, this method is highly unscalable. If all routers
operate according to Algorithm \ref{alg:flood}, then \erase\
messages are guaranteed to be delivered to every  $R \in \mathbb{R}_N$. However,
the number of routers receiving a specific \erase\ message is much
larger than $|\mathbb{R}_N|$. Therefore, flooding should always be
the last resort for \erase\ messages. We assess the actual overhead of this
technique in Section \ref{sec:simulations}.

\subsection{Forwarding History Analysis}
We now analyze performance of lossless and lossy forwarding
histories described in Section \ref{sec:fw_histories}.

\subsubsection{Lossless Histories}
The memory (and possibly computational) cost of a lossless forwarder history
grows as a function of $t$. Thus, history collection will inevitably
saturate memory at some point. Let $n_{max}^R$ be the total size
(in entries) of the history memory for $R$. Saturation is reached at time $t$ such that
$n_t^R \geq n_{max}^R$. We compute the time required to saturate
a lossless forwarder history in two scenarios. We assume that each content
object is $4,096$ bytes and hash digests are $32$ bytes.

\begin{compactitem}
\item {\bf Consumer-facing router:} Assume a caching consumer-facing router
(e.g., an access point) with $4$GB of history storage and
data rate of $100$ Mbps. This data rate is equivalent to a
content forwarding rate of $\mu_F^R = 3'200$ Cps (content per second). If the
router operates at full capacity with a full cache -- i.e., storing every forwarded
content requires the eviction of an already cached one --  it will take $41,943$ secs.
for history storage to be saturated. This is roughly $12$ hours.
This window of time might be longer than the {\tt ExpiryTime} of
content objects that are subject to be erased. For instance, news feed pages
are likely to be updated with a frequency faster than $1/12$ hours.
\item {\bf Core router:} Assume a non-caching CCN core router with $1$TB of flash
history storage and data rate of $10$ Tbps, i.e., equivalent to $\mu_F^R = 335$
MCps. Assuming such a router is always working at full capacity (always forwarding
$10$ Tbps), the lossless forwarder history can be saturated in $102$ secs. In this
case producers have a time window of less than $2$ minutes to issue an \erase\
message for content $C$ after it was last served.
\end{compactitem}
$R$'s saturation time can be lengthened by increasing the available size of the
forwarder history. However, at this rate, the cost of adding more memory
to make the saturation time useful is far too expensive ($1$TB for $2$ minutes of history).

A very natural question arises: what happens when $R$'s history storage is saturated?
$R$ can evict old history entries randomly or according to some policy,
e.g., LRU. However, keeping track of history entries' ages
might lead to reduced performance. Another alternative
is to divide history storage into smaller chunks, each
corresponding to a set time window of history entries. Once history storage is saturated,
the oldest chunk is erased to provide space for new entries. Using the consuming-facing
router example above, $4$GB of history storage can be divided into $12$ chunks, each corresponding
to one hour. The router could then erase the history recorded $12$ hours ago in order to
store history entries for the coming hour.

\subsubsection{Lossy Histories}
Lossy histories are useful when lossless histories are too expensive, e.g., in
core network routers. Our approach to lossy forwarder history is
based on Bloom Filters (BFs) -- probabilistic data structures with tunable performance.
Given an $m$-bit BF that stores
$n$ elements, the number of input hash functions $k$ can be optimized and false
positive probability can be estimated using Equation \ref{equ:bloom_fpp_opt}
\cite{broder2004network}. Note that optimal value of $k$ is also given as a
function of $m$ an $n$.
\begin{align}
\label{equ:bloom_fpp_opt}
f(m, \cdot, n) &\approx \left( 0.6185 \right)^\frac{m}{n}, \quad k = \ln(2) \cdot \frac{m}{n}
\end{align}
In practice, a router can optimizes the number of hash functions in order to
lower false positive probability. An upper bound of $k$ can be set to
limit hashing overhead.

As mentioned above, standard BFs do not support entry deletion, which is
necessary to deal with the saturation problem. As indicated in \cite{snoeren2001hash},
historical information for Internet-scale traffic (IP packets)
can not last beyond a few minutes, which might still be less than what
we needed for BEAD.

We now analyze lossy forwarding history in the context of two scenarios mentioned
above with the same history storage and data rates. We also assume
that each content object added to 4BF changes the value of new distinct $k$
bits from $0$ to $1$. Clearly, this is unrealistic, since we do not consider the
possibility of overlapping of hash function outputs for different input elements.
However, this assumption captures the worst-case scenario.
\begin{compactitem}
\item {\bf Consumer-facing router:} To maintain a maximum false positive probability
of $10^{-32}$, a BF of size $4$GB can fit $n \leq 2 \times 10^8$ elements.
Based on Equation \ref{equ:bloom_fpp_opt}, it requires $k = 120$ hash
functions. Thus, it will take $89'478$ secs. (a little over one day) for
forwarder history to be saturated.
\item {\bf Core router:} To maintain the same false positive probability, a
BF of size $1$TB can accommodate $n \leq 5.7 \times 10^8$ elements, which corresponds
to $k = 107$ hash functions. Forwarding history will
be saturated in $245$ secs.
\end{compactitem}
One major drawback to using BFs for lossy forwarding histories is that history
saturation is more difficult to resolve. Recall that, with lossless histories,
a router can remove old entries in order to add new ones.
A router could also delete the oldest chunk of the history once it is saturated.
However, with lossy histories, a router can either: (1) flush the entire lossy history and
start over, or (2) use CBFs which support element deletion with the use of counters.
Unfortunately, this introduces false negative probabilities.

\subsubsection{Packet Marking Analysis}
Packet marking  is computationally inexpensive
since it requires a single MAC computation per (either interest or
\erase) packet. However, its drawback is increased memory footprint of the
interest along every hop. Recall that traces in the hop-sequence consist of: (1) router
identifier, (2) interface identifier, and (3) tag. Assuming a $2$-byte interface identifier
and a SHA-256-based MAC, the total size of each trace is $38$ bytes. This corresponds to
extra $608$ bytes for each interest, assuming a 16-hop router-level path.\footnote{The average Internet
hop-count is currently $16$ \cite{begtasevic2001measurements}.}

We now compare two hop-sequence techniques described in Section
\ref{sec:interest_marking}. Assume a tree topology with (1) one producer $P$ at the
root with height $h$, (2) $2^h$ consumers at the leaves with height $0$, and (3)
$2^h-2$ routers. We assume all consumers request content $C$ and all routers append
hop-sequence traces to the corresponding interests. In this case, $P$ receives
$2^h$ interests, each with $h-1$ traces. If $P$ includes all these traces
in a single \erase\ message, its size would be
grow by $\left(2^h~\cdot~(h-1)\right)\times 38$ bytes. This becomes $35$MB
for $h=16$, which is clearly impractical.\footnote{We defer designing
a more efficient scheme for combining hop-sequence traces to future work.} On the
other hand, if $P$ decides to send a separate \erase\ to each consumer it
would generate $2^h$ \erase\ messages. The same overall volume of traces ($35$MB)
will be sent from $P$ to consumers. However, it would be split into numerous \erase\ messages.
One advantage is that \erase\ messages size will likely not exceed the path
MTU and therefore not require fragmentation.

\subsection{Summary of the BEAD Protocol}
Various approaches for routing \erase\ messages are practical in
different network locations. For instance, consumer-facing (caching)
routers can keep lossless or lossy histories for
at least a day. Meanwhile, interest marking is better for core network routers.
Therefore, we believe that all aforementioned techniques can be used in combination
for routing \erase\ messages. Our recommendations are:
\begin{compactenum}
\item\label{itm:marking} If $R$ supports interest marking and the first tuple
in the hop-sequence traces is valid and appended by the router itself, then
information in the tuple is used to route the \erase\ downstream.
\item\label{itm:cache} If the  content is in $R$'s cache, then the in-cache history
is used to route the \erase.
\item\label{itm:histories} If the content is not in $R$'s cache, but $R$ keeps
lossless or lossy histories, then they are used for \erase\ message routing.
\item\label{itm:flooding} Otherwise, $R$ floods received
\erase\ messages according to Algorithm \ref{alg:flood}.
\end{compactenum}
Recommendation \ref{itm:marking} is most appropriate for core network
routers, \ref{itm:cache} and \ref{itm:histories} for less busy edge network routers, and
\ref{itm:flooding} as a failover mechanism. Most routers
would likely prefer to drop \erase\ messages instead of flooding them. This is why BEAD
is a \emph{best-effort} protocol: it does not guarantee that each \erase\ message will
be delivered to all entities caching the target content.

As mentioned before, not all published content is subject to
future deletion. If routers can make this distinction,
then there is no need to record history entries about content that will not be deleted.
Such distinction can be achieved by adding an optional {\tt CanERASE} flag to content object
headers. If this flag is not present, the default behavior is to assume that no \erase\ messages
will be sent for the corresponding content. Moreover, interests requesting content that will
not be deleted are not required to be marked by routers. Producers could
tell consumers what content is subject to deletion (i.e., an \erase) by overloading
catalogs or manifests. As described in \cite{ghali2014network} and \cite{kurihara2015encryption},
catalogs and manifests contain lists of Self-Certifying Names (SCNs) of content
to be requested. This list is provided by the producer and can contain the {\tt CanERASE} flag
alongside each SCN. In this case, the interest header format should be modified to include this
optional flag. Since it is not guaranteed that all content objects will be requested using SCNs,
i.e., catalogs, the default behavior of (core) routers is to append hop-sequence traces to
interests if the {\tt CanERASE} flag is missing.

%% file: 05-experiments.tex
\begin{figure*}[t]
\begin{center}
\subfigure[DFN topology.]
{
	\includegraphics[width=0.7\columnwidth]{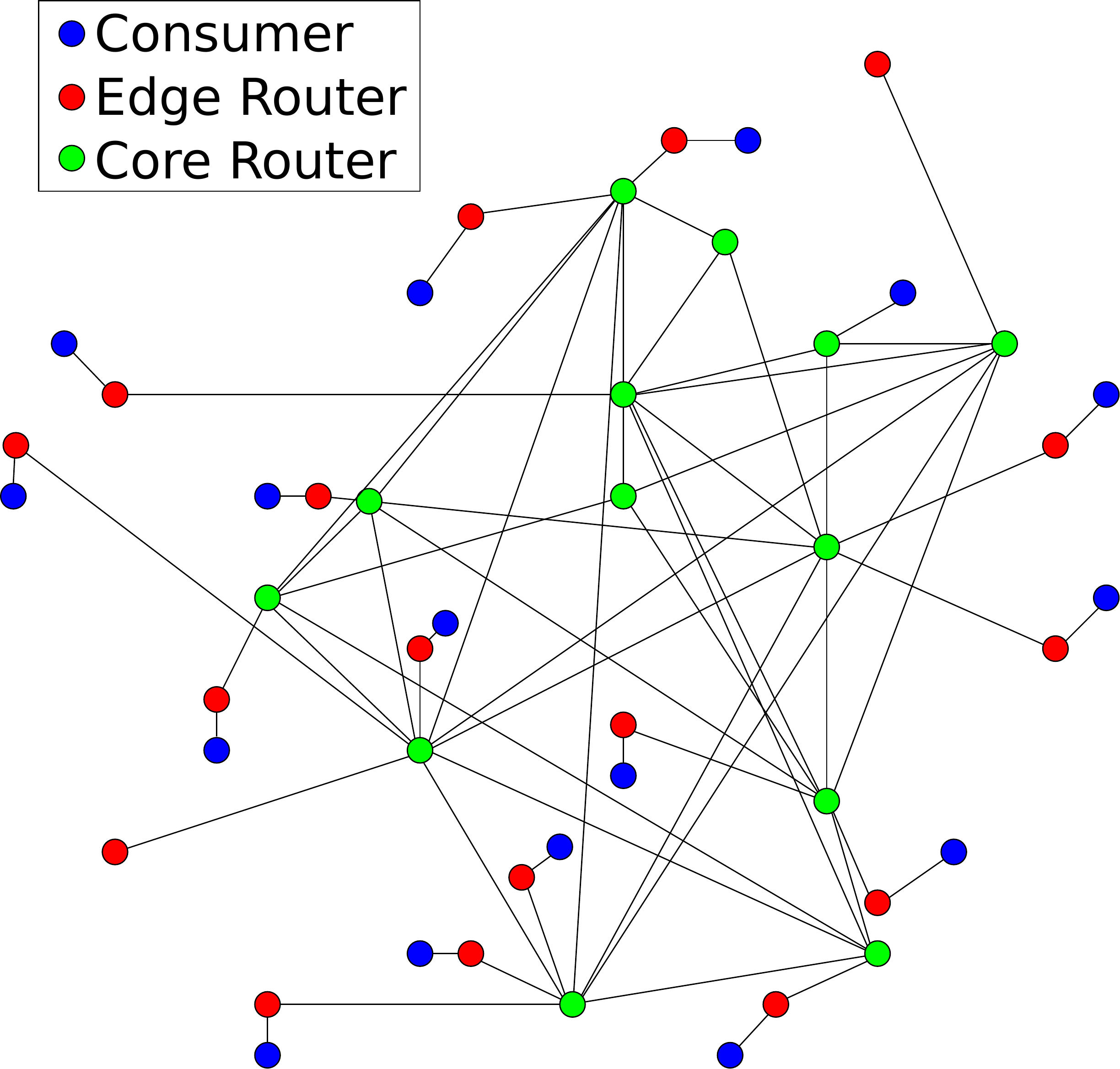}
	\label{fig:dfn-topology}
}
\subfigure[AT\&T topology.]
{
	\includegraphics[width=\columnwidth]{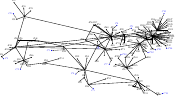}
	\label{fig:att-topology}
}
\caption{The DFN and AT\&T topologies.}
\label{fig:tops}
\end{center}
\end{figure*}

\begin{figure*}[t]
\begin{center}
\subfigure[Data processing overhead in the DFN topology with $160$ consumers.]
{
	\includegraphics[width=\columnwidth]{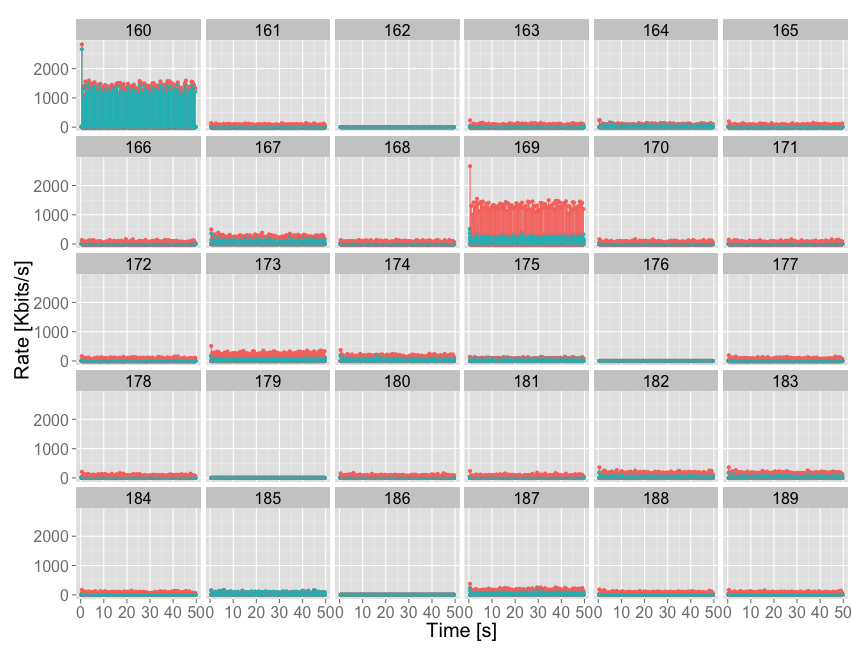}
	\label{fig:dfn-1}
}
\subfigure[\erase\ messages processing overhead in the DFN topology with $160$ consumers.]
{
	\includegraphics[width=\columnwidth]{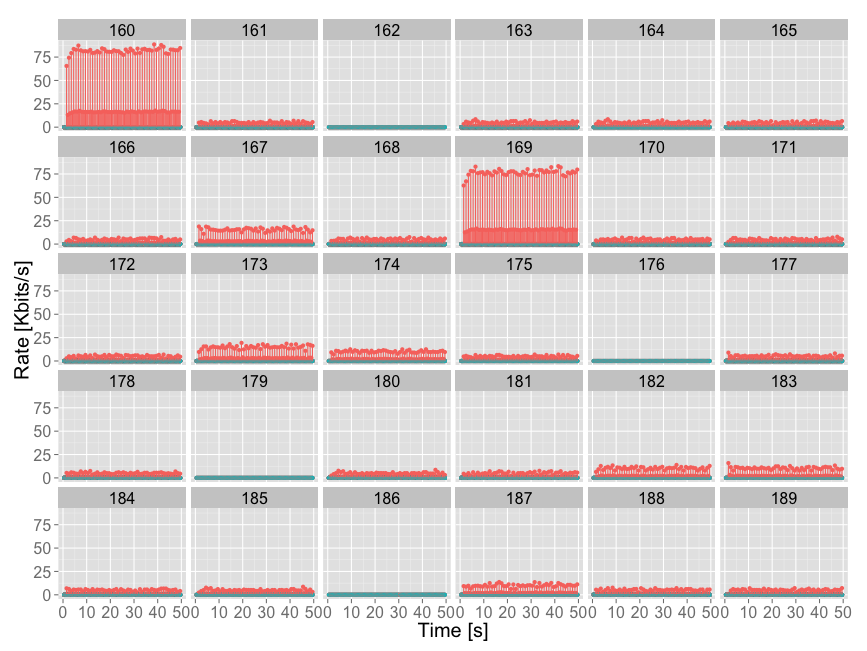}
	\label{fig:dfn-2}
}
\subfigure[Data processing overhead in the AT\&T topology with $160$ consumers.]
{
	\includegraphics[width=\columnwidth]{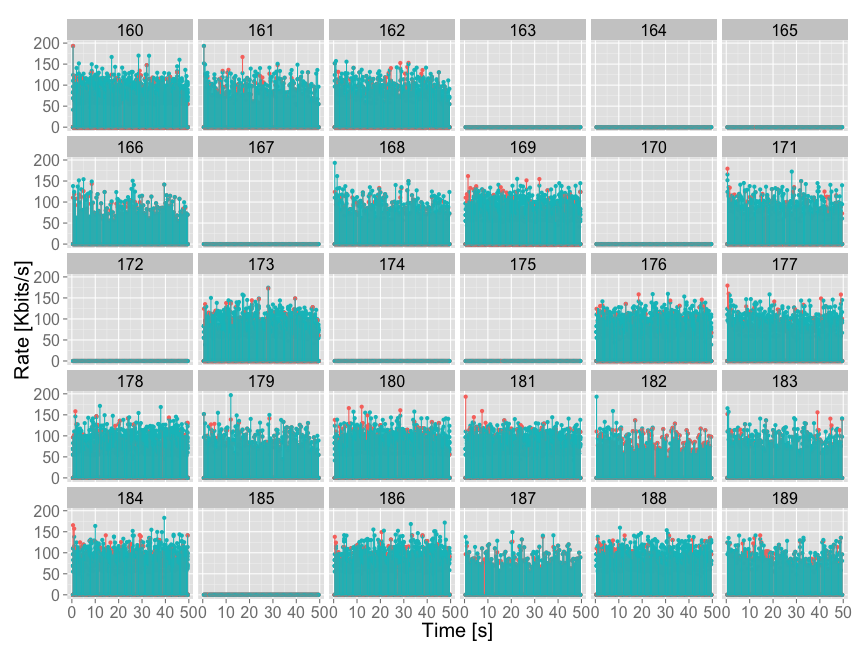}
	\label{fig:att-1}
}
\subfigure[\erase\ messages processing overhead in the AT\&T topology with $160$ consumers.]
{
	\includegraphics[width=\columnwidth]{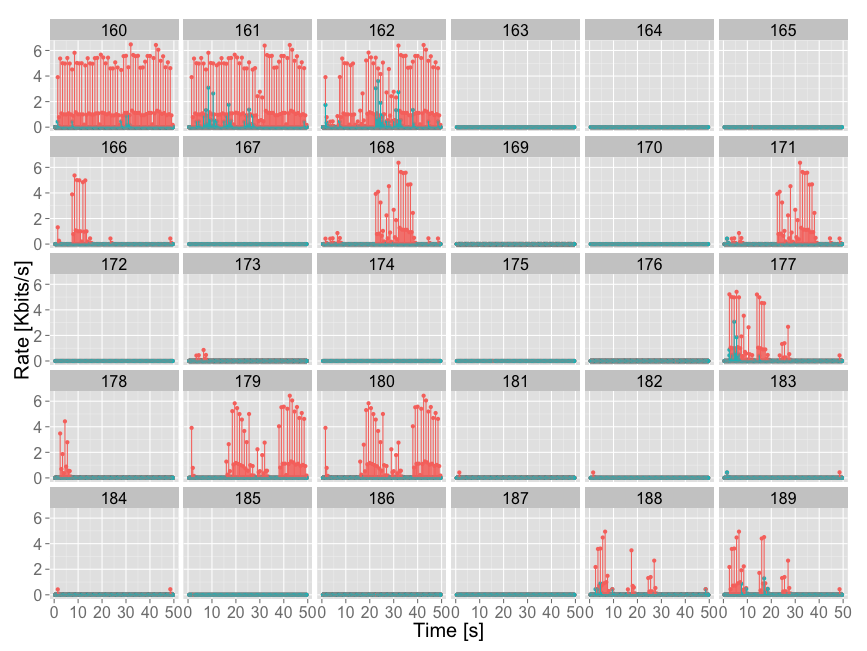}
	\label{fig:att-2}
}
\caption{Network overhead from processing \erase\ messages. Routers are identified
by integers in the range [160..189]. InData (OutData) and InErase (OutErase) correspond
to the amount of content object and \erase\ traffic received from (sent to) an
upstream (downstream) node, respectively. Ingress data is shown in red and egress
data is shown in blue.}
\label{fig:bead-results}
\end{center}
\end{figure*}

\begin{figure*}[t]
\begin{center}
\subfigure[DFN topology with $160$ consumers.]
{
	\includegraphics[scale=0.26]{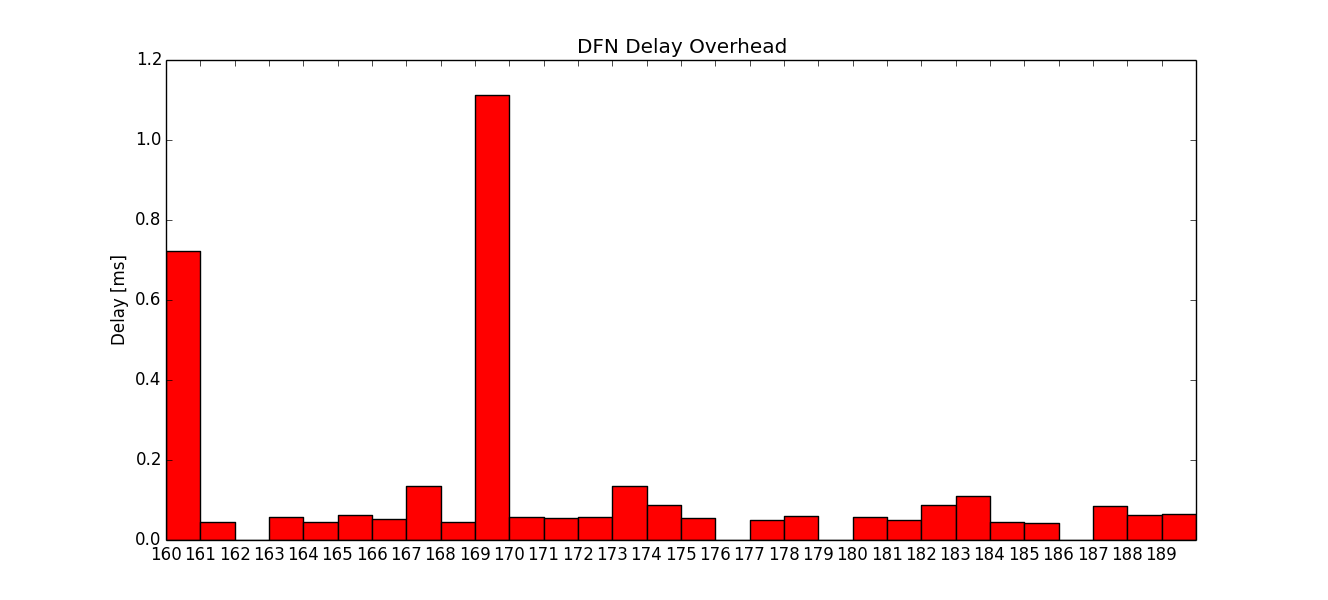}
	\label{fig:dfn-overhead}
}
\subfigure[AT\&T topology with $160$ consumers.]
{
	\includegraphics[scale=0.26]{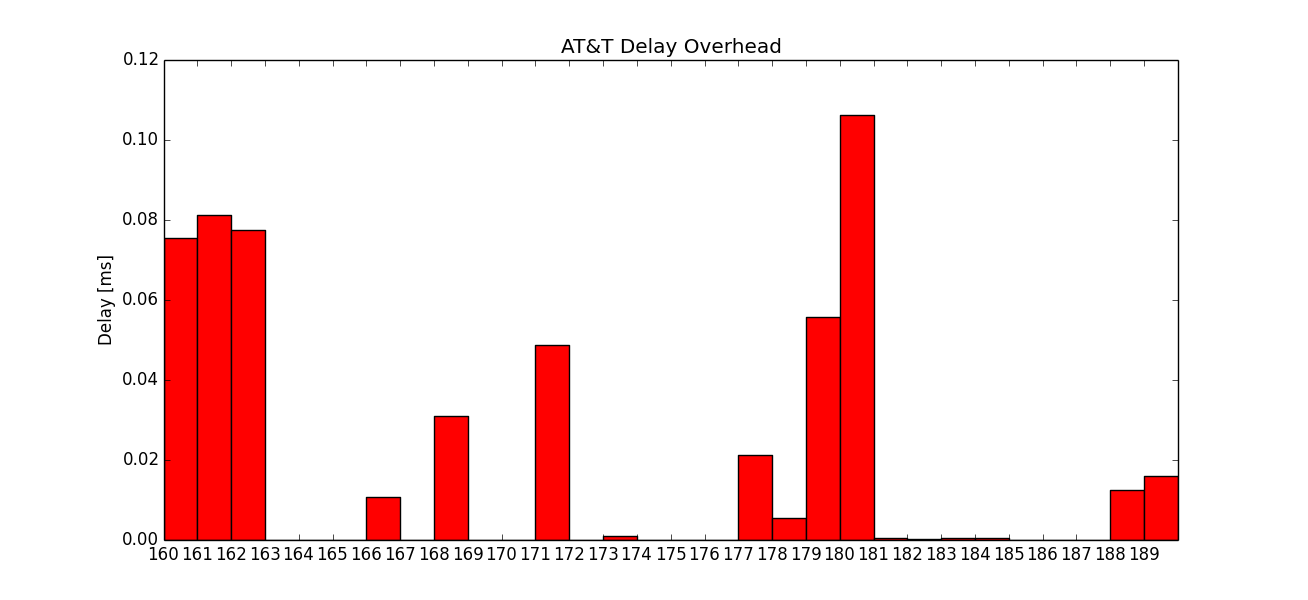}
	\label{fig:att-overhead}
}
\caption{Forward \erase\ processing overhead in the DFN and AT\&T topologies.
The results are captured for each of the routes assessed in the bandwidth
overhead experiments. Routers are identified by integers in the range [160..189]
and correspond to the routers in Figure \ref{fig:bead-results}.}
\label{fig:overhead}
\end{center}
\end{figure*}

\section{Experimental Simulations} \label{sec:simulations}
Our experimental simulations are focused on two properties of BEAD: network
overhead (in terms of additional bytes added for \erase\ messages) and
forwarder overhead for processing \erase\ messages, i.e., the average amount
of time it takes to process each \erase.

\subsection{Network Overhead}
To measure the network overhead due to generating and forwarding \erase\
messages we extended ndnSIM 2.0 \cite{ndnsim} -- an implementation of
NDN architecture as a NS-3 \cite{ns3} module for simulation purposes -- to
support \erase\ messages. With this modified architecture, we ran two sets of
experiments using the following topologies (shown in Figure \ref{fig:tops}):
\begin{compactitem}
\item The DFN network, Deutsches ForschungsNetz (German Research Network)
\cite{DFNverein, DFN-NOC}: a German network developed for research and
education purposes which consists of $30$ connected routers positioned
in different areas of Germany. The blue dots in the figure represent group
of consumers ($10$ consumers per blue dot) connected to edge routers (red dots),
while the green dots represent core network routers.
\item The AT\&T backbone network~\cite{compagno2013poseidon}. This consists of
over 130 routers. Each logical consumer in the figure represents multiple ($5$)
physical consumers connected to an edge router.
\end{compactitem}
In all experiments, consumers issue requests at a rate of $10$ interests per second
for content with the name prefix \path{/prefix/A} and monotonically increasing
sequence number suffix. Every router uses a lossless history to record previously
forwarded content objects for \erase\ forwarding. Lastly, producers
issue \erase\ messages for 50\% of their content every 1 second. Under these conditions,
we measure router packet processing overhead with respect to content objects and
\erase\ messages. Figures \ref{fig:dfn-1} and \ref{fig:dfn-2} compare the overhead of processing
content objects and \erase\ messages in the DFN topology with $160$ consumers. Similarly,
Figures \ref{fig:att-1} and \ref{fig:att-2} show the same type of overhead in the AT\&T
topology with the same number of consumers. Comparatively, we find that
\erase\ messages contribute very little overhead to the network with respect to the
bandwidth consumed by content objects. Specifically, the total amount of \erase\
messages traffic in the DFN topology is $1.8\%$ of the total content objects traffic,
whereas it is only $0.09\%$ in the AT\&T topology.

We also assessed the actual computational
overhead incurred by each router in these scenarios. The average time to process
a single \erase\ message for the DFN and AT\&T scenarios are shown in Figures
\ref{fig:dfn-overhead} and \ref{fig:att-overhead}. We see that only a subset
of the routers incur greater than 1.0ms to process an \erase. These routers
are those closest to the producer since they almost always receive, store, and
forward \erase\ messages.

%% file: 06-discussion.tex
\section{Monetizing Content Deletion} \label{sec:monotizing}
We now discuss potential economic incentives for routers and ISPs to
support content deletion and implement the BEAD protocol.

\subsection{BEAD with Accounting}
So far, we discussed how the network routes \erase\ messages towards
routers that possibly cache corresponding content. The main challenge
is that producers do not know where such content is cached. We also acknowledge
that BEAD is a best-effort protocol, unless flooding is used, which is undesirable.

However, if producers knew exactly where content is cached, then
\erase\ messages could be routed efficiently. For example, if a producer knew that
a particular AS had a copy of the content cached \emph{by some node in the system},
then the producer could specifically ask the AS to distribute an \erase\
internally. This is far superior to routing \erase\ messages in the core of the network
in hopes that they {\em might} reach this AS (and any others with a cached copy).
%%GTS: the next few lines are useless and obvious. Please tell me if you really think they add
% something?
%%CHRIS: without further explanation the reader will think, "how are they routed efficiently?"
% we need to be clear how routing would be "targeted" to somewhere instead of "hopeful"
%%%CESAR: in the accounting paper (at least the published one), we don't introduce
% management servers, that's why I think explaining it here is important.

We believe that it is possible to distribute content caching location information along with accounting information.
In a recent secure CCN accounting scheme \cite{ghali2015practical},
Ghali et al. propose that routers should notify producers of content they serve from caches
by sending a so-called ``push interest'' or \pint. This approach can be modified such that:
(1) AS gateways send \pints\ when content is cached in their domain
and (2) \pint\ messages carry the prefix of an AS accounting management 
server within the AS.\footnote{Accounting management servers are centralized
entities that manage accounting activities inside the AS.}
Whenever a producer wants to delete certain content, it sends
an \erase\ message to each accounting management server (one per AS) that previously
reported caching corresponding content. Then, the latter distribute the \erase\
message within their ASs. Intra-AS distribution can be achieved via techniques described in 
Section \ref{sec:routing}. In fact, flooding might well be appropriate for 
that purpose since erase messages would not traverse AS boundaries.

The relationship between accounting and BEAD is natural. This is because 
one of the important applications of accounting is to bill for
cache space. From an economic perspective, it would not be surprising for in-network
caching to become a paid service. Routers and ASs could offer caching services
for producers. A reasonable extension to this service would be to also offer a
deletion service via BEAD.

\subsection{BEAD in the Core}
Flooding in the network core is not a viable as a means of distributing
\erase\ messages. Moreover, forwarder histories and packet marking are (relatively)
expensive operations and too costly for the fast path in the core.
ISPs will likely just drop these messages due to a lack of economic
incentive to forward them. Thus, in any plausible CCN network
-- where producers and consumers are at the edges of a network, while most
traffic is routed through the core -- \erase\ messages are most likely to be propagated along
only half of producer-to-consumer path(s). This is troublesome since content is
most likely to be cached near consumers in edge (or near-edge) routers, and
\erase\ messages might never reach these routers.

To address this issue, core routers must be incentivized to carry and
forward \erase\ messages from producers to consumers. Since \erase\ messages will
typically amplify traffic, producers should be expected to pay for this increase. 
As before, this effectively turns BEAD into a service provided by ISPs that 
complements monetized caching; producers who pay for cache space may also have 
the choice to pay for on-demand deletion via BEAD.

%% file: 08-conclusion.tex
\section{Conclusion} \label{sec:conclusion}
We proposed BEAD -- a best-effort autonomous deletion protocol. BEAD is designed
to solve the problem of stale or unsafe content in CCN. We described an efficient
and lightweight form of authenticator for BEAD deletion requests and discussed
several ways in which they could be routed from producers to consumers. We assessed
the performance of each technique and verified the network overhead and deletion
``penetration'' using simulations. For future work, we will formalize the integration
of accounting and BEAD to form a comprehensive platform for monetized caching
in CCN.